\documentclass[aps,pra,preprint,showpacs,showkeys,floatfix]{revtex4-1}
\usepackage[pdftex]{graphicx}
\usepackage{bm}
\usepackage{color}
\usepackage{amssymb}
\usepackage{bbold}

\renewcommand{\deg}{$^{\circ}$\hspace{1mm}}

\newcommand{\etal}{ {\it et al.}}
\newcommand{\newc}{\newcommand}
 
\newc{\be}{\begin{equation}}
\newc{\ee}{\end{equation}}
\newc{\bfe}{\begin{floatequation}}
\newc{\efe}{\end{floatequation}}
\newc{\bea}{\begin{eqnarray}}
\newc{\eea}{\end{eqnarray}}
\newc{\ie}{{\it i.e.} }
\newc{\eg}{{\it e.g.} }
\newc{\etc}{{\it etc.} }
\newc{\ra}{\rightarrow}
\newc{\lra}{\leftrightarrow}
\newc{\lsim}{\buildrel\langle\over{\sim}}
\newc{\gsim}{\buildrel\rangle\over{\sim}}

\newc{\one}{\mathbbm{1}}
\newc{\Tr}[1]{\mathrm{Tr}\left[ {#1} \right]}
\newc{\ket}[1]{\left|{#1}\right\rangle}
\newc{\bra}[1]{\left\langle{#1}\right|}
\newc{\braket}[2]{\langle{#1}|{#2}\rangle}
\newc{\mean}[1]{\langle{#1}\rangle}
\newc{\braketd}[1]{\langle{#1}|{#1}\rangle}
\newc{\ketbrad}[1]{\left|{#1}\rangle\!\langle{#1}\right|}
\newc{\ketbra}[2]{\left|{#1}\rangle\!\langle{#2}\right|}
\newc{\EV}[2]{\langle{#1}\rangle_{#2}}
\newc{\C}{\ensuremath{\mathbbm C}}

\def\be{\begin{equation}}
\def\ee{\end{equation}}

\begin{document}

\title{Optimization of Measurement Device Independent Scarani-Ac\`{i}n-Ribordy-Gisin protocol}

\author{C. Tannous\footnote{Tel.: (33) 2.98.01.62.28,  E-mail: tannous@univ-brest.fr} and J. Langlois} 
\affiliation{Laboratoire des Sciences et Techniques de l'Information, de la Communication et 
de la Connaissance, UMR-6285 CNRS, Brest Cedex3, FRANCE}

\begin{abstract}
The measurement device independent (MDI) Quantum Key Distribution (QKD) 
is a practically implementable method for transmitting secret keys 
between respective partners performing quantum communication.  
SARG04 (Scarani-Ac\`{i}n-Ribordy-Gisin 2004) is a protocol
tailored to struggle against photon number splitting (PNS) attacks by
eavesdroppers and its MDI-QKD version is reviewed and optimized from secret key 
bitrate versus communication distance point of view.
We consider the effect of several important factors such as error correction function, dark 
counting parameter and quantum efficiency in order to achieve the largest 
key bitrate versus longest communication distance.
 \end{abstract}



\maketitle

\date{\textcolor{blue}{Version \today}}

While Classical cryptography use two types of keys to
encode and decode messages (secret or symmetric and public or asymmetric keys)
Quantum cryptography uses QKD for transmitting secret keys 
between partners allowing them to encrypt and decrypt their messages.
QKD principal characteristic is that it is practically implementable and has already 
been deployed  commercially by several quantum communication providers such as SeQureNet in France, 
IQ Quantique in Switzerland, MagiQ Technologies in the USA and QuintessenceLabs in Australia.
The second main feature of QKD is that it allows communicating parties to
detect online eavesdroppers in a straightforward fashion.

In principle, QKD is unconditionally secure nevertheless its practical implementation
has many loopholes and consequently has been attacked by many different ways 
exploiting some intermediate operation or another during secret key processing such as
time-shift~\cite{attack4,attack5}, phase-remapping~\cite{attack6}, 
detector blinding~\cite{attack7,attack8}, detector dead-time~\cite{attack9}, 
device calibration~\cite{attack10}, laser damage~\cite{attack11}...

This work is about optimization of SARG04~\cite{Scarani} MDI-QKD version protocol designed
to fend off photon number splitting (PNS) attacks by considering important factors such as
error correction function types, detector dark counting parameter and quantum efficiency. 
It is organized as follows: after reviewing the original four-state SARG04 protocol,
we discuss its MDI version and describe the effects of various parameters
on communication distance and secret key bitrate.

SARG04 protocol has been developed to combat PNS attacks that are targeted
toward intercepting photons present in weak coherent pulses (WCP) that are used for
communication. This stems from the fact, it is not possible presently to commercially  exploit 
single photons in a pulse. However, progress in developing large scale methods targeted at 
using single photons in a pulse is advancing steadily.

SARG04 being very similar to BB84~\cite{Scarani} protocol, the simplest example 
of secret key sharing among sender and receiver (Alice and Bob), we review first the BB84 case below.

In the BB84 protocol framework, Alice and Bob use two channels to communicate: one quantum and private to send
polarized single photons and another one classical and public (telephone or Internet) 
to send ordinary messages~\cite{Tannous}.
Alice selects two bases in 2D Hilbert space consisting each of two
orthogonal states: $\bigoplus$ basis with $(0,\pi/2)$
linearly polarized photons,   
and $\bigotimes$ basis with $(\pi/4, -\pi/4)$ linearly polarized photons.

Four photon polarization states: $\ket{\rightarrow}, \ket{\uparrow},\ket{\nearrow},\ket{\searrow}$
are used to transmit quantum data with
$\ket{\nearrow}=\frac{1}{\sqrt{2}}(\ket{\rightarrow}+ \ket{\uparrow})$
and $\ket{\searrow}=\frac{1}{\sqrt{2}}(\ket{\rightarrow}- \ket{\uparrow})$.

A message transmitted by Alice to Bob over the Quantum channel is a stream of 
symbols selected randomly among the four above and Alice and Bob choose randomly 
one of the two bases $\bigoplus$ or $\bigotimes$
to perform photon polarization measurement.

Alice and Bob announce their respective
choice of bases over the public channel without revealing the measurement results.

The raw key is obtained by a process called "sifting" consisting of retaining only the 
results obtained when the used bases for measurement  are same.

After key sifting, another process called key distillation~\cite{Scarani} must be performed.
This process entails three steps~\cite{Scarani}: error correction, privacy amplification and
authentication in order to counter any information leakage from
photon interception, eavesdropping detection (with the no-cloning theorem~\cite{Scarani}) and
exploitation of announcement over the public channel.

The basic four-state SARG04 protocol is similar to BB84 but adds a number of steps to
improve it and protect it against PNS attacks. The steps entail introducing 
random rotation  and filtering of the quantum states. Before we describe it, we introduce 
some states and operators~\cite{Yin} using Pauli matrices $\sigma_X,\sigma_Y,\sigma_Z$:

\begin{itemize}

\item $R=\cos(\frac{\pi}{4})I-i\sin(\frac{\pi}{4})\sigma_Y$ is a $\pi/2$ rotation operator about $Y$ axis,

\item $T_0=I$ is the (2$\times$2) identity operator, 

\item $T_1=\cos(\frac{\pi}{4})I-i\sin(\frac{\pi}{4})\frac{(\sigma_Z+\sigma_X)}{\sqrt{2}}$ is a $\pi/2$ rotation operator around the $(Z+X)$ axis,

\item $T_2=\cos(\frac{\pi}{4})I-i\sin(\frac{\pi}{4})\frac{(\sigma_Z-\sigma_X)}{\sqrt{2}}$ is a $\pi/2$ rotation operator around the $(Z-X)$ axis. 

\end{itemize}

Alice prepares many pairs of qubits and sends each one of them to Bob after performing a random
rotation over different axes with $T_{l}R^k$  where $l\in\{0,1,2\}$ and $k\in\{0,1,2,3\}$.

Upon receiving the qubits, Bob first applies:

\begin{itemize}
\item  A random reverse multi-axis rotation $ R^{-k'}T_{l'}^{-1}$, 

\item Afterwards, he performs a local filtering operation defined by  
$F=\sin(\frac{\pi}{8})\ket{0_x}\bra{0_x}+\cos(\frac{\pi}{8})\ket{1_x}\bra{1_x}$ where
$\{\ket{0_x},\ket{1_x}\}$ are $X$-eigenstate qubits; 
they are also eigenvectors of $\sigma_X$ with eigenvalues +1, and -1 respectively.
Local filtering enhances entanglement degree and the $\pi/8$ angle helps retrieve~\cite{Tamaki} 
one of the maximally entangled EPR Bell~\cite{EPR,Kwiat} states i.e. polarization 
entangled photon pair states given by:
$\ket{\psi^\pm}=\frac{1}{\sqrt{2}}(\ket{\rightarrow \uparrow} \pm \ket{\uparrow \rightarrow}),
\ket{\phi^\pm}=\frac{1}{\sqrt{2}}(\ket{\rightarrow \rightarrow} \pm \ket{\uparrow \uparrow})
$. They form a complete orthonormal basis in 4D Hilbert space for all polarization states
of a two-photon system and the advantage of local filtering is to make Alice and Bob 
share pairs of a Bell state making the shared bits unconditionally secure~\cite{Tamaki}.

\item  After, Alice and Bob compare their indices ${k,l}$ and ${k',l'}$ via public communication, and keep the qubit pairs with $k=k'$ and $l=l'$ when Bob's filtering operation is successful.

\item They choose some states randomly as test bits, measure them in the $Z$ basis, and compare their results publicly to estimate the bit error rate and the information acquired by the eavesdropper.

\item Finally, they utilize the corresponding Calderbank-Shor-Steane (CSS) code~\cite{CSS} to correct bit and phase errors and perform a final measurement in the $Z$ basis on their qubits to obtain the secret key.

\end{itemize}

Following Lo \etal~\cite{Lo2012} Mizutani \etal~\cite{Mizutani} modified the original 
SARG04 protocol by including an intermediate experimental setup run by Charlie, 
at mid-distance between Alice and Bob, 
consisting of Bell correlation measurements. The setup contains a half beam-splitter, 
two polarization beam-splitters to simulate photonic Hadamard and CNOT gates in order to
produce Bell states, as well as  photodiode detectors. This additional step will help 
discard non perfectly anti-correlated photons and thus reduce transmission error rates.
In addition, Alice and Bob not only choose photon polarization randomly, they also
use WCP amplitude modulation to generate decoy states in order to confuse the eavesdropper. 

The protocol runs as follows:

\begin{itemize}

\item Charlie  performs Bell measurement on the incoming photon pulses 
and announces to Alice and Bob over the public channel 
whether his measurement outcome is successful or not. 
When the outcome is successful, he announces the successful events  
as being of Type1 or Type2. Type1 is coincidence detection events 
of $AT$ and $BR$ or $BT$ and $AR$. Type2 is coincidence detection events 
of $AT$ and $AR$ or $BT$ and $BR$ where $AT,BT$ stand for detecting transmitted $(T)$ photon events
from Alice $(A)$ or Bob $(B)$ linearly polarized at 45\deg whereas $AR, BR$ 
are for detecting reflected $(R)$ photon events at -45\deg.

\item  Alice and Bob broadcast $k$ and $k'$, 
over the public channel. 
If the measurement outcome is successful with Type1 and $k=k'=0,\ldots,3$, 
they keep their initial bit values, and Alice flips her bit. 
If the measurement outcome is successful with Type2 
and $k=k'=0, 2$, they keep their initial bit values. 
In all the other cases, they discard their bit values.

\item   After repeating the above operations several times, 
Alice and Bob perform error correction, privacy amplification 
and authentication as described previously.

\end{itemize}

In the ideal case (no transmission errors, no eavesdropping) 
Alice and Bob should discard results pertaining to 
measurements done in different bases (or when Bob failed to detect 
any photon).  

\begin{figure}[htbp]
  \centering
    \resizebox{80mm}{!}{\includegraphics[angle=0,clip=]{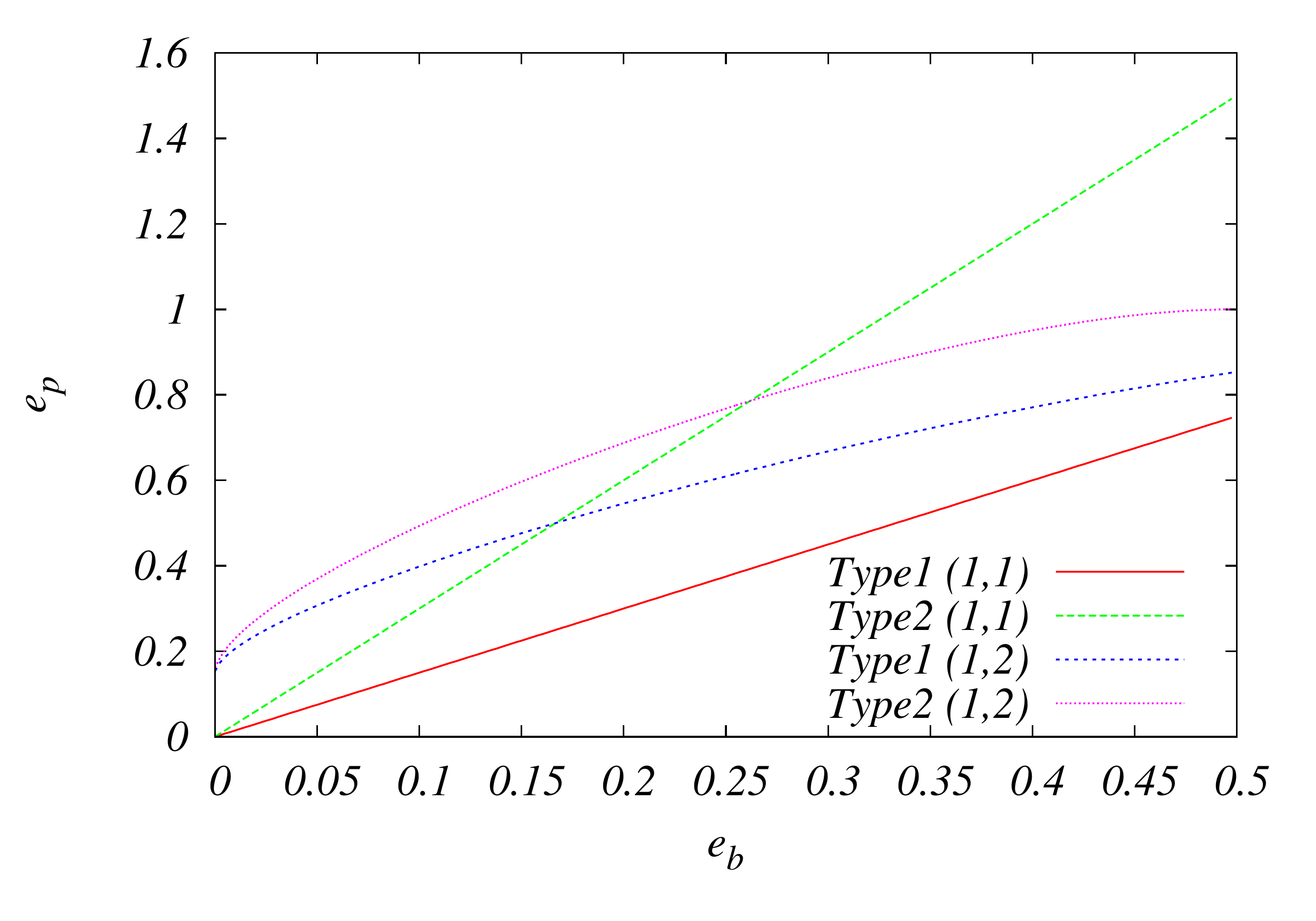}}  
\vspace*{-3mm} 
\caption{(Color on-line) Phase error probability $e_p$ versus bit error probability 
$e_b$ for Type1 and Type2 with different number of photons $(m,n)$ emitted by Alice and Bob.}
\label{epeb}
\end{figure}  

In QKD, Alice and Bob should be able to determine efficiently their shared secret key 
as a function of distance $L$ separating them. Since, the secure key is determined after
sifting and distillation, secure key rate is expressed in bps (bits per signal) given
that Alice sends symbols to Bob to sift and distill with the remaining bits making the secret key.

For Type $i$ event, we define $e_{i,p}^{(m,n)}$ as the phase error probability
that Alice and Bob emits $m$ and $n$ photons respectively, 
and Charlie announces a successful outcome with $Q_i^{(m,n)}$, 
the joint probability. Consequently the asymptotic key rate for Type $i$ is given as a sum
over partial private amplification terms of the form $Q_i^{(m,n)}[1-h_2(e^{(m,n)}_{i,{p}})]$
and one error correction term $Q_i^{tot}f(e_i^{tot})h_2(e_i^{tot})$ related to total
errors as~\cite{Mizutani,GLLP}: 
\bea
K_i(L) &=Q_i^{(1,1)}[1-h_2(e^{(1,1)}_{i,{p}})]+Q_i^{(1,2)}[1-h_2(e^{(1,2)}_{i,{p}})] \nonumber\\
&+Q_i^{(2,1)}[1-h_2(e^{(2,1)}_{i,{p}})]-Q_i^{tot}f(e_i^{tot})h_2(e_i^{tot}). 
\label{keyrate}
\eea

The total probabilities $Q_i^{tot}=\sum_{m,n}Q_i^{(m,n)}$ 
and total error rates are given by $e_i^{tot}=\sum_{m,n}Q_i^{(m,n)}e^{(m,n)}_{i,{b}}/Q_i^{tot}$
where $e^{(m,n)}_{i,{b}}$ is the Type $i$ bit error probability and
$h_2$ is the binary Shannon entropy~\cite{Carlson} given by $h_2(x)=-x\log_2(x)-(1-x)\log_2(1-x)$. 
Moreover, the above asymptotic key rate is obtained in the limit of infinite number of 
decoy states~\cite{Mizutani}.

Phase error probabilities are determined from bit error probabilities as depicted in fig.~\ref{epeb}
for Type 1 and 2 and depending on photons $(m,n)$ emitted.

Since Charlie is in the middle between Alice and Bob, 
the channel transmittance to Charlie from Alice is the same as that from Bob. 
Considering that $L$ is the distance between Alice and Bob, 
the channel transmittance $\eta_T$ is obtained by replacing $L$ by $L/2$ resulting in: 
$\eta_T=10^{-\alpha{L/20}}$.

For the standard Telecom wavelength~\cite{Carlson}
$\lambda=1.55 \mu$m, the loss coefficient with distance is $\alpha$=0.21 dB/km.
The quantum efficiency and the dark count rate of the detectors are taken as 
$\eta=0.045$ and $d=8.5\times 10^{-7}$, respectively as in the GYS~\cite{GYS} case.

We compare below the effect of a fixed error correction function with respect to a fixed
value function.

The error correction function is given by Enzer \etal~\cite{Enzer} as:
$ f_e(x)=1.1581+57.200 x^3$

\begin{figure}[htbp]
  \centering
    \resizebox{80mm}{!}{\includegraphics[angle=0,clip=]{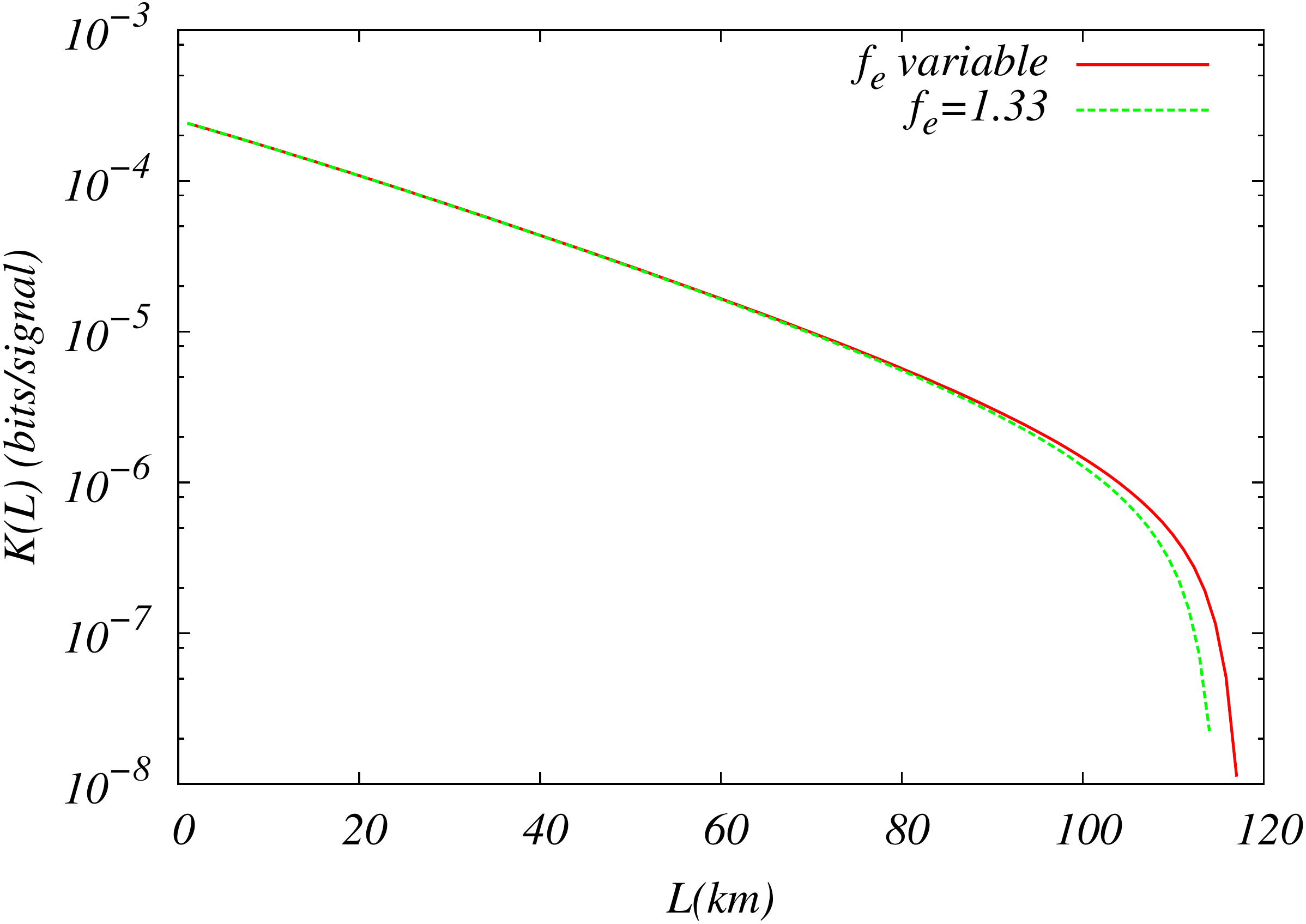}}  
\vspace*{-3mm} 
\caption{(Color on-line) Key rate $K(L)$ for Type 1 events, in bps versus distance $L$ using the same parameters 
as in Ref.~\cite{GYS} $\eta$=0.045, $d=8.5 \times 10^{-7}$, $\alpha=0.21$ for different error correction function.}
\label{Rate1}
\end{figure}

\begin{figure}[htbp]
  \centering
    \resizebox{80mm}{!}{\includegraphics[angle=0,clip=]{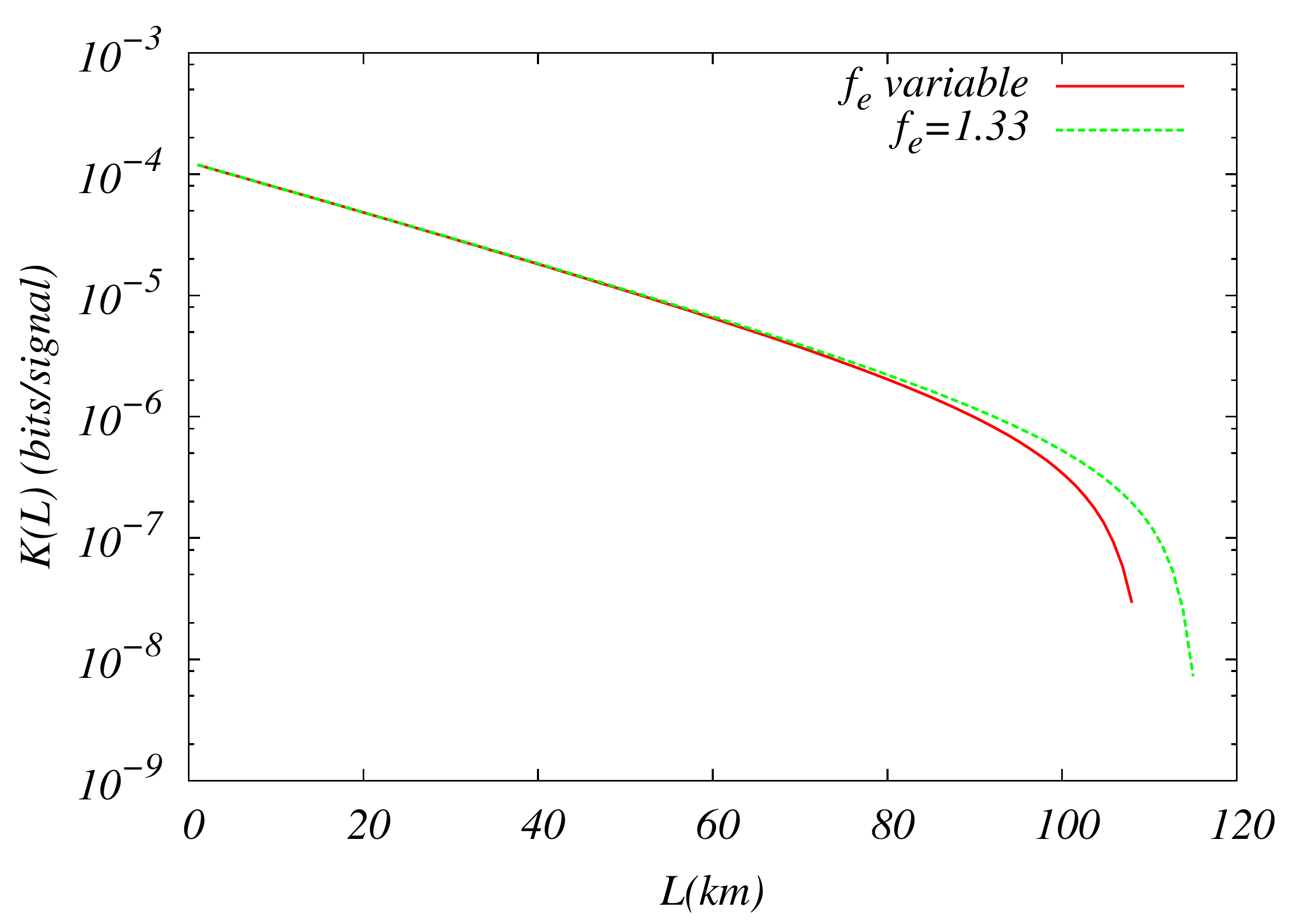}}  
\vspace*{-3mm} 
\caption{(Color on-line) Key rate $K(L)$ for Type 2 events, in bps versus distance $L$ using the same parameters 
as in Ref.~\cite{GYS} $\eta$=0.045, $d=8.5 \times 10^{-7}$, $\alpha=0.21$  for different error correction function.}
\label{Rate2}
\end{figure}

In figs~\ref{Rate1},\ref{Rate2} secret key rates for Type 1 and Type 2 events 
are displayed versus distance when $f_e$ function is
considered as variable or fixed at a value of 1.33.

Improving quality of detection means that dark counting must be substantially reduced
in order to avoid false "clicks" (irrelevant event detection) of the detectors.

\begin{figure}[htbp]
  \centering
    \resizebox{80mm}{!}{\includegraphics[angle=0,clip=]{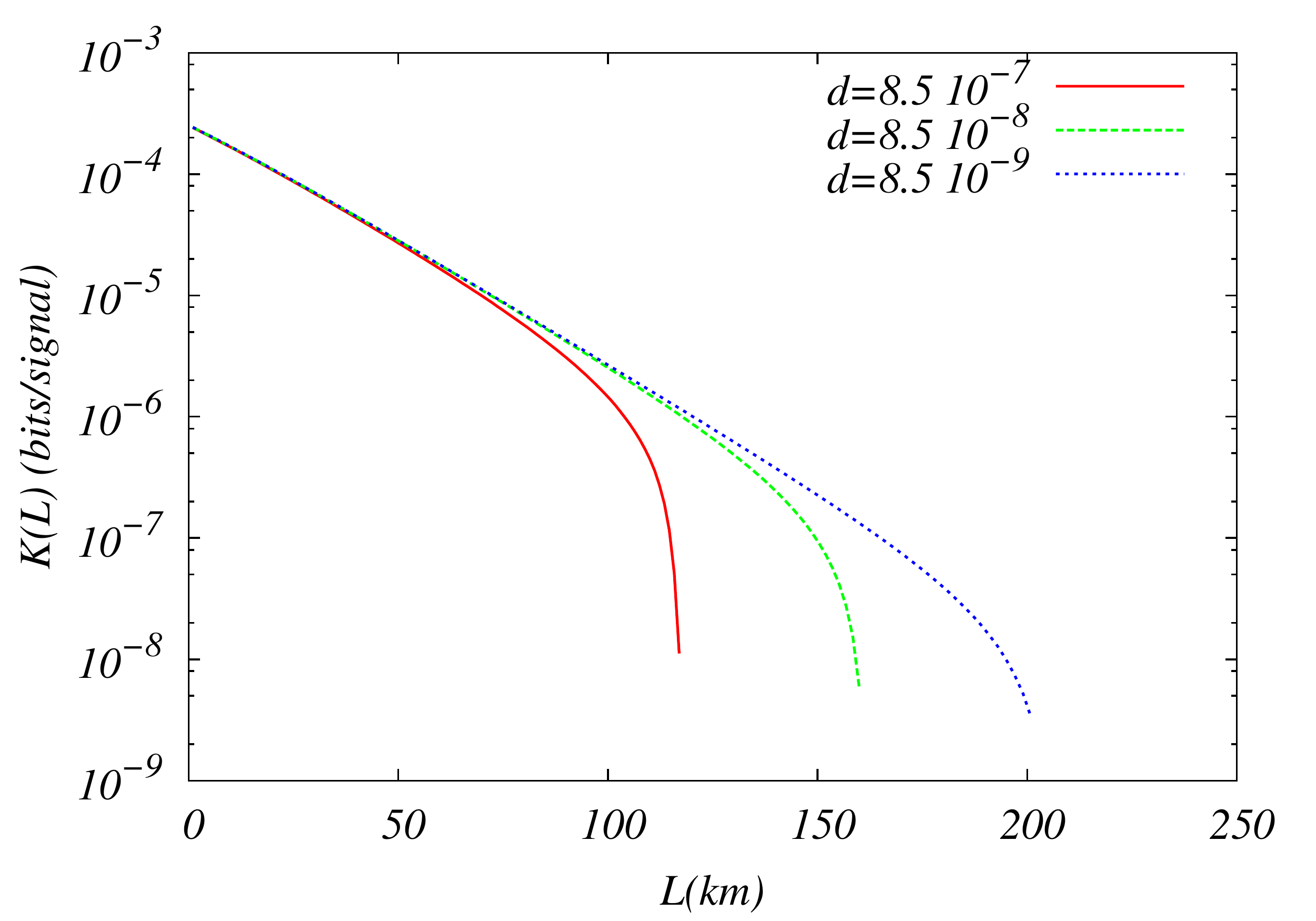}}  
\vspace*{-3mm} 
\caption{(Color on-line) Key rate $K(L)$ for Type 1 events, in bps versus distance $L$ using the same parameters 
as in Ref.~\cite{GYS} $\eta$=0.045, $\alpha=0.21$  for different detector dark counting parameter $d$ and variable error correction function.}
\label{Rate1_d}
\end{figure}

\begin{figure}[htbp]
  \centering
    \resizebox{80mm}{!}{\includegraphics[angle=0,clip=]{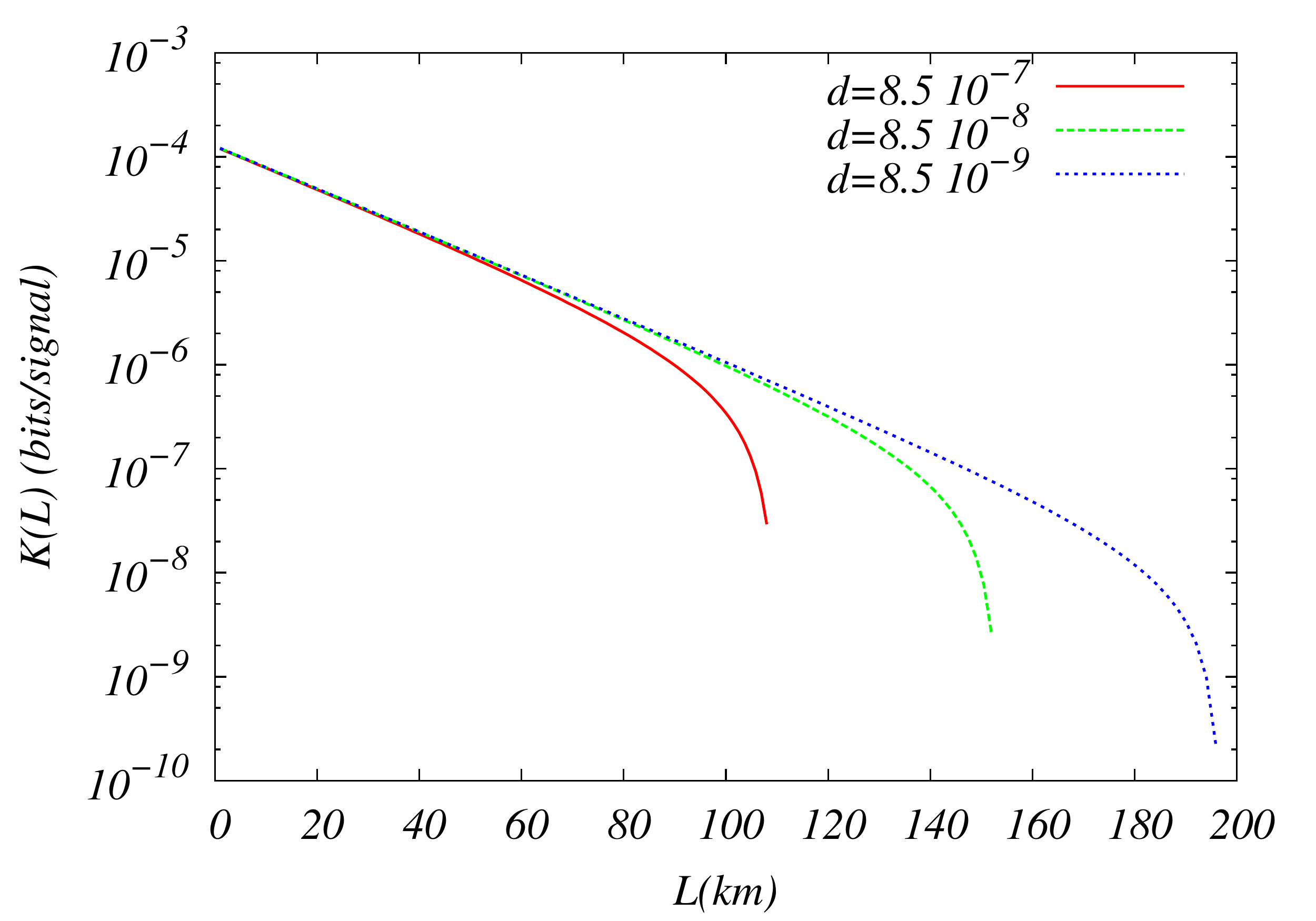}}  
\vspace*{-3mm} 
\caption{(Color on-line) Key rate $K(L)$ for Type 2 events, in bps versus distance $L$ using the same parameters 
as in Ref.~\cite{GYS} $\eta$=0.045, $\alpha=0.21$  for different detector dark counting parameter $d$ and variable error correction function.}
\label{Rate2_d}
\end{figure}

In figs~\ref{Rate1_d},\ref{Rate2_d} secret key rates for Type 1 and Type 2 events 
are displayed versus distance for different values of the dark count rate with error correction
function $f_e$ freely varying.

%---------------------update--------------------------------------------------

Quantum yield is an important parameter that plays an important role in quantum
communications.

\begin{figure}[htbp]
  \centering
    \resizebox{80mm}{!}{\includegraphics[angle=0,clip=]{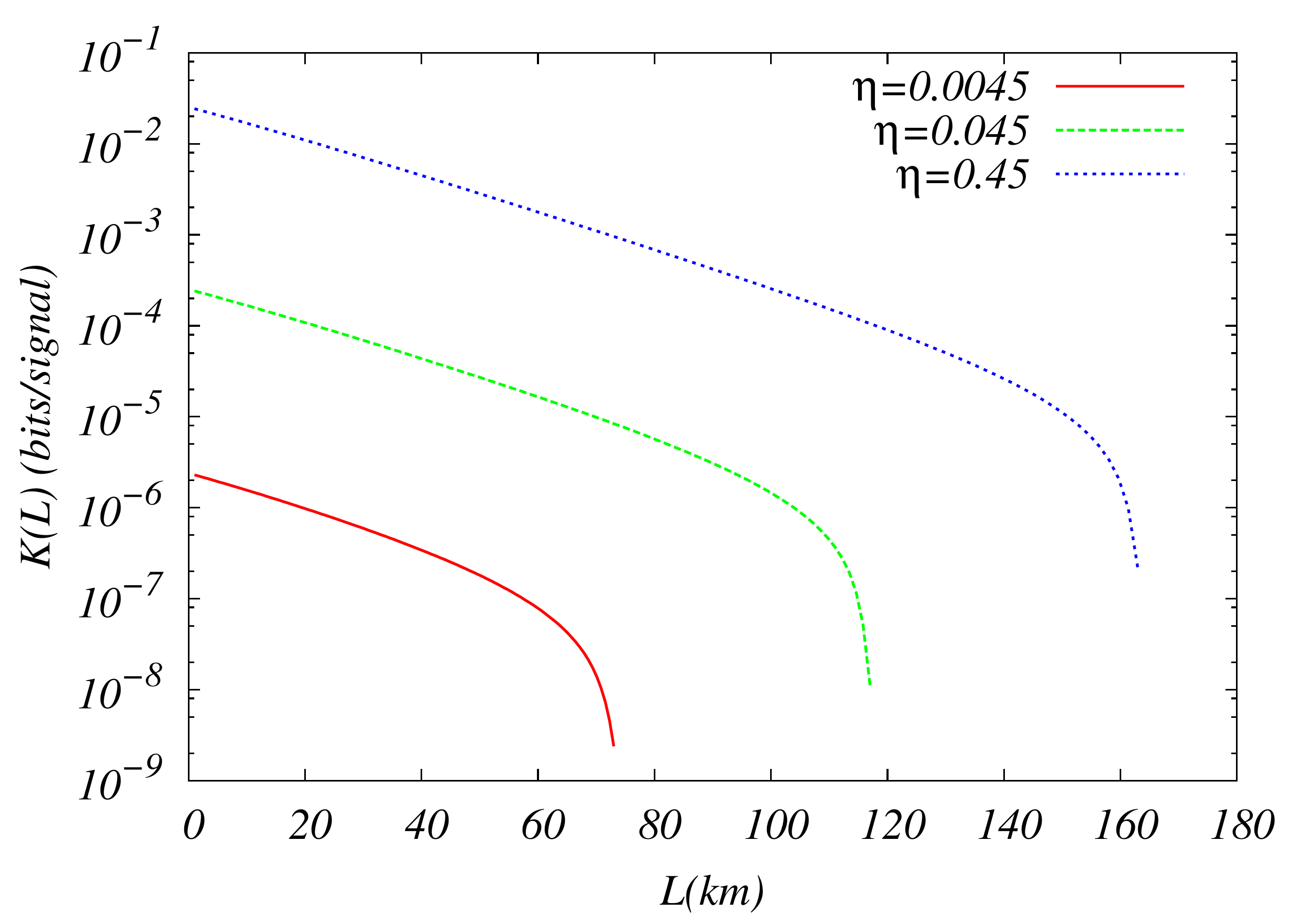}}  
\vspace*{-3mm} 
\caption{(Color on-line) Key rate $K(L)$ for Type 1 events, in bps versus distance $L$ using the same parameters 
as in Ref.~\cite{GYS} $d=8.5 \times 10^{-7}$, $\alpha=0.21$ and $f_e$ variable for different values of
the quantum yield parameter $\eta$.}
\label{Rate1_Y}
\end{figure}

\begin{figure}[htbp]
  \centering
    \resizebox{80mm}{!}{\includegraphics[angle=0,clip=]{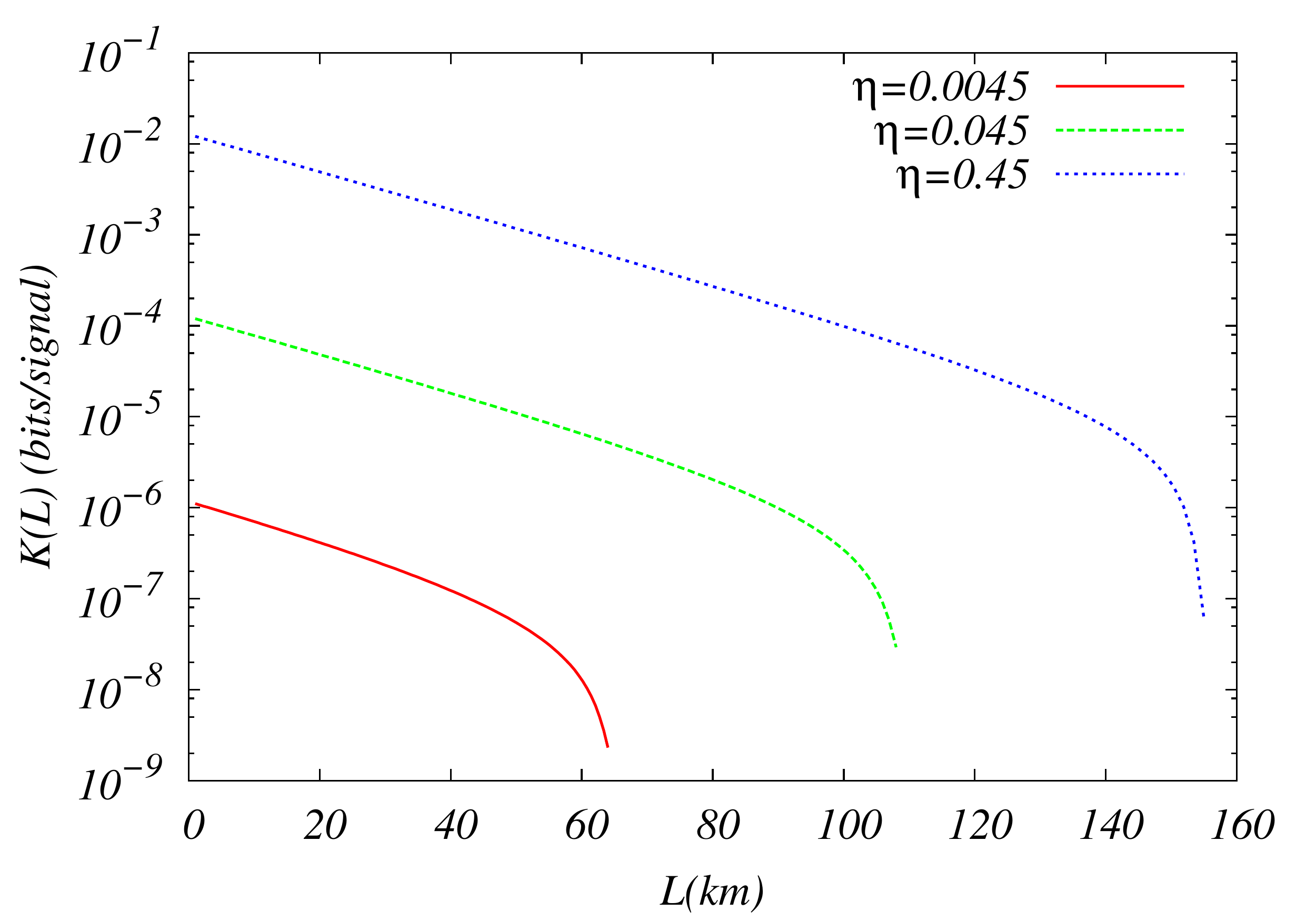}}  
\vspace*{-3mm} 
\caption{(Color on-line) Key rate $K(L)$ for Type 2 events, in bps versus distance $L$ using the same parameters 
as in Ref.~\cite{GYS} $d=8.5 \times 10^{-7}$, $\alpha=0.21$ and $f_e$ variable for different values of
the quantum yield parameter $\eta$.}
\label{Rate2_Y}
\end{figure}

In figs~\ref{Rate1_Y},\ref{Rate2_Y} secret key rates for Type 1 and Type 2 events 
are displayed versus distance for different values of the quantum yield $\eta$ with error correction
function $f_e$ freely varying. The value of $\eta$ has been intentionally exaggerated in order to
explore the range of communication distances covered by it variation. It is interesting to note that
the Quantum yields acts on communication distance and key bitrate simultaneously whereas dark count
rate and error correction function changes affect solely communication distance.

Communication distances and secret key bitrates obtained in this work can be improved
when we vary the error correction function, dark count rate and quantum efficiency. 
Insight into SARG04 protocol acquired by optimization leads to conclude that the most sensitive way to increase communication distance substantially is to decrease the dark count rate.
The least sensitive parameter is the error correction function type and in spite of exaggerating
the values of the quantum efficiency in order to probe the largest possible range of communication
distances, the dark count rate parameter is the most promising. Consequently future research efforts ought to be directed towards reducing it considerably. This improvement relies on developing special algorithms  
that will allow to discriminate between different events occurring around the photodetectors or 
developing materials with selective and specially engineered higher thresholds preventing false 
"clicks" triggered by "irrelevant" events.

\end{document}